\documentclass[aps,prl,groupedaddress]{revtex4}
\usepackage{graphicx} 
\begin{document}
\title{No 'cut off' in the High Energy Cosmic Ray Energy Spectrum}

\author{Tadeusz Wibig}
\affiliation{Experimental Physics Dept., University of \L \'{o}d\'{z}
}
\affiliation{ The Andrzej So\l tan Institute For Nuclear Studies,
Cosmic Ray Lab., \L \'{o}d\'{z}, Uniwersytecka 5, POB 447, \L \'{o}d\'{z} 1 ;Poland}
\author{Arnold W. Wolfendale}
\affiliation{Physics Department, University of Durham, Durham , UK.}

\begin{abstract}
It is often claimed that there should be a 'GZK cut-off" in the flux of extragalactic 
cosmic rays, arising from interactions between the cosmic rays and the cosmic micro-wave 
background photons (e.g. \cite{1} and \cite{2}).  Some experiments (\cite{3} and \cite{4}) show particles of even 
higher energy than this value and this has led to claims for exotic processes 
(e.g. \cite{5} and \cite{6}).

We contend that such claims are unnecessary -  there is no predicted cut-off, rather a 
continuation of the injection spectrum at reduced intensity. We have combined the world's 
data and shown that the prediction for a rather flat injection spectrum (exponent: 1.9 - 2.2) 
in the case of universal particle injection provides a reasonable fit to the data.  
Conventional forms for the particle attenuation in the intergalactic medium (e.g. \cite{7} 
and \cite{8}) have been assumed.  Either protons or iron nuclei (or a mixture) will suffice.

Attention is drawn to another aspect, too, that of the losses on the infra-red radiation 
which may be intense near to strong sources and for sources in galaxy clusters.  The 
attendant magnetic fields near the sources leads to significantly long diffusion times 
through the strong infra-red fields.  Two 'case histories' are considered.
\end{abstract}
\maketitle

The question of the origin of cosmic rays in general is a difficult one, that of those 
of the highest energies (known to mankind) is singularly so.  Here, we examine the 
situation above about 10$^{18}$ eV (the spectrum extends to beyond 10$^{20}$~eV) in the light 
of very recent knowledge.  Our aims are three fold:  to correct misapprehension about 
the  so-called 'Greisen-Zatsepin-Kuzmin cut-off', to consider the evidence for a 
transition from Galactic (G) to extragalactic (EG) origin near 10$^{19}$ eV and to stress 
the need to consider the strong infra-red background radiation (IRB) near very strong 
sources (such as quasars) and sources in galaxy clusters, which are often considered 
as prime candidates for ultra-high-energy cosmic ray sources (UHECR).

Following our earlier work (Ref.\cite{9}) we have updated our combination of the world's data.  
We find that by normalizing energies and intensities to the position of the minimum 
in the (usual) plot of log E$^3\times$I(E) vs. logE we are able to secure a reasonably 
consistent set of data from the 6 sets of results considered (the arrays at Volcano 
Ranch, Haverah Park, AGASA, (\& AKENO), Yakutsk, Hi-Res and Fly's Eye - see \cite{9} for 
references) up to about  3$\times$10$^{19}$ eV.  At higher energies there is a significant 
disparity between measurements, with conventional terrestrial particle arrays, 
typified by AKENO, and those involving use of the Cherenkov or Fluorescence technique, 
typified by the 'Hi-Res' array.  The displacements in energy involved are typically 0.2 
in logE, i.e. not much higher than the random errors.

We proceed in alternative directions.  The first case (Figure 1(a)) assumes that the Hi-Res
data have accurate energy and intensity calibrations; the second case - which can be regarded 
as an upper limit in energy calibration - assumes that the AGASA calibration is correct (Figure 1(b)).

Several observations can be made about the Figures, as follows:
\begin{itemize}
\item[(i)]
The data below log E = 19.5 (Case (a)) and 19.8 (Case (b)) show 
an excellent fit to the sum of two components: G (Galactic) and EG 
(extragalactic).  G falls with increasing energy because of a lack 
of sufficiently energetic sources and a lack of magnetic trapping. 
(see Ref.\cite{10}) and the EG spectrum initially is a power law before 
losses on the CMB set in.  We would expect, a priori, that the EG 
spectrum would have an exponent in the region 1.8 to 2.4, this being 
the injection spectrum expected for Fermi-style acceleration.  Some 
explanation of the range is necessary, as follows.  It has been shown 
that the standard value of 2.0 is not necessarily the minimum value - 
harder spectra can be generated by the first order mechanism in shocks 
propagating into plasmas with low beta values \cite{11}. Softer spectra, i.e. 
exponents bigger than 2.0 appear where losses are important during the 
acceleration process.  It is well known that below log E = 15, Galactic 
sources have an injection spectrum in this region - the steeper spectrum 
observed at low energies is because of the energy-dependent Galactic 
trapping caused by an energy-dependent diffusion coefficient for escape.  
We regard the excellence of the fit of the points to the sum of G and EG 
as a strong indication that the transition from G to EG occurs here 
(logE~=~18.7 case (a), 18.9 case (b)).  We find no support for the 
arguments \cite{12} favouring a transition at a much lower energy (log~E~=~17.4 - a factor 20 lower in energy).  It would be difficult to 
achieve such a sharp 'ankle' if all the particles in the region were extragalactic.

 The Extragalactic energy density for our case (a) is 10$^{-6}$~eV~cm$^{-3}$.
\item[(ii)]
Fits have been made for commonly considered cases, where the sources are 
distributed smoothly throughout the universe, with one proviso - we take 6 Mpc 
as the minimum distance to a source.  The reason for this proviso is that 
closer sources would almost certainly be seen as such.  We \cite{14} claimed 
some years ago a tentative sighting of a nearby pair of colliding galaxies 
at 7 Mpc.  The claim has not (yet) been confirmed but colliding galaxies 
are still considered a possible mechanism for UHECR origin.  Insofar as 
the mean density of 'normal' galaxies is about 3$\times$10$^{-2}$~Mpc$^{-3}$, 
the number of galaxies within a radius of 6 Mpc is  $\sim$ 9, including our own.

The whole philosophy of considering 'normal' galaxies is to imagine that 
they contain Galaxy-like sources which extend to more than 10$^{20}$ eV 
but that, by chance, our Galaxy does not contain one, or more, at the 
present time \cite{13}.  More specifically, UHECR from Galactic sources are 
not arriving at the present time.

\item[(iii)]
 The prediction of Ref.\cite{1}, denoted `T' in Fig 1 (b), which shows a rapid 
monotonic 'cut-off' has been much quoted and, indeed, used as a claim 
for new physics for the case where particles are observed above 10$^{20}$ eV.  
We consider this steep line to be inappropriate.

\item[(iv)]
Our prediction using standard attenuation factors for interactions with 
the radiation fields (principally the CMB) present universally, are shown in each case.
\end{itemize}

Insofar as we normalize the attenuation factors at log E = 19 there is not much difference 
between the predicted spectra for proton - and iron - primaries.  There will be some 
difference at lower energies but uncertainties in the shape of the Galactic spectrum 
allow these differences to be compensated.

It is evident that if the injection spectrum is flat enough (differential exponent $\gamma$
 = 2.2 for case (a) and 1.9 for case (b)) a tolerable fit can be achieved to the data. 
The peak in case (a) can perhaps be accounted for in terms of a stronger than average 
infra-red background (IRB) near the sources; this topic is taken up in more detail later.

What is certain at this stage is that there is no 'GZK cut-off' expected, for physically 
reasonable injection spectra.  It can be added, in parenthesis, that the term 'cut-off' 
was a mis-translation from the original Russian (V.Kuzmin, private communication).

We turn now to EG 'systems' which are much stronger in radio, optical, X-rays and 
gamma rays than normal galaxies, these potential UHECR sources include active 
galactic nuclei (AGN), quasars, galaxy clusters and, as remarked already, 
colliding galaxies.  Our particular concern is to draw attention to the fact 
that they will be surrounded by strong IR fields and regions of significant 
magnetic fields associated with low energy CR escaping from the AGN.  
Interestingly, the IRAS infra-red satellite showed \cite{15} that many AGN 
sources were stronger in IR than in all other wavelengths and that 
colliding galaxies (mergers) are often associated with very strong 
infra-red emission.  We can take a strong quasar as an example; 
with a luminosity LIRB $\sim$ 2$\times$10$^{45}$~erg~s$^{-1}$ 
and it produces an IRB of  $\sim$ 5$\times$10$^{-2}$~eV~cm$^{-3}$ 
at a distance of 1 Mpc.  There will be significant spectral distortion 
for particles (protons or nuclei) escaping from such a source due to 
interactions with the enhanced IRB.  Concerning quasars, a problem 
appears concerning their distance.  Whereas the range of a 2$\times 10^{20}$~eV 
proton is $\sim$ 55 Mpc against CMB interactions, most quasars are further afield.
   Particles of lower energy will not arrive either within the Hubble Time because 
of the magnetic field in the IGM (a few nG): the diffusion time is too long.

Distant quasars are important, however, if - as seems likely - they are an 
important source of UHECR in the Universe.  The interaction of their particles 
with the CMB and the enhanced IRB will produce gamma rays and neutrinos which 
have the possibility of arrival and detection at earth.  Calculations so far 
made may have underestimated the likely intensity of these secondary particles.

It is with perturbations to the spectral shape of UHECR from more local sources 
that we are more concerned here.  Such sources are galaxies in general - see the 
previous discussion - AGN, colliding galaxies and galaxies in clusters.  The 
effect of enhanced IRB in galaxies is negligible when averaged over all sources
 ($\sim$0.1\%).  That in colliding galaxies is more serious insofar as many of 
the most prominent IR emitters are, in fact, galaxies in collision \cite{15}.  
Equally serious will be the situation for strong sources in galaxy clusters.  
The nearest cluster is VIRGO (at 15 Mpc) and this contains the celebrated AGN:
 M87.  There is also the likelihood of strong shocks in clusters giving rise 
to UHECR.

Figure 2(a) gives an indication of the expected magnitude of the effect of 
losses (IR and CMB) on protons leaving a source surrounded by a magnetic 
field and IRB of its own making.

Figure 2(b) gives results for more modest sources in a cluster of galaxies.
  (AGN, shocks within the intercluster medium, etc.).Measurements have shown 
that many clusters have magnetic fields in the region of 5~$\mu$G \cite{16}
and calculations have been made with B = 1~$\mu$G and 5~$\mu$G.  The IRB has 
been taken as in Figure 2(a), viz with the IRB energy density at R = 0.3 Mpc 
increasing by 100 (the IGM value) at R = 3 Mpc and higher closer in still.  
It is evident that the reduction in the important energy region near 10$^{19}$~eV 
can be large; this factor may lead to some suppression of the bump in figure 1 
predicted for a uniform.  EG source distribution, such as cluster-contained 
sources would be, in first order.  Clearly, accurate calculations for specific 
source models will need to take losses by way of infra-red radiation seriously.
 
We conclude that there is no GZK 'cut-off' expected in the energy spectrum of 
UHECR if the injection spectrum is sufficiently flat and extends far enough 
in energy.  The problem is 'how do UHECR get their extremely high energies 
in the first place' rather than 'how is it that such particles are able to 
reach us?' Concerning the effect of infra-red radiation losses, these can 
be serious for very strong sources and for strong sources in galaxy clusters 
containing significant magnetic fields.

\centerline{
\includegraphics[width=12.cm]{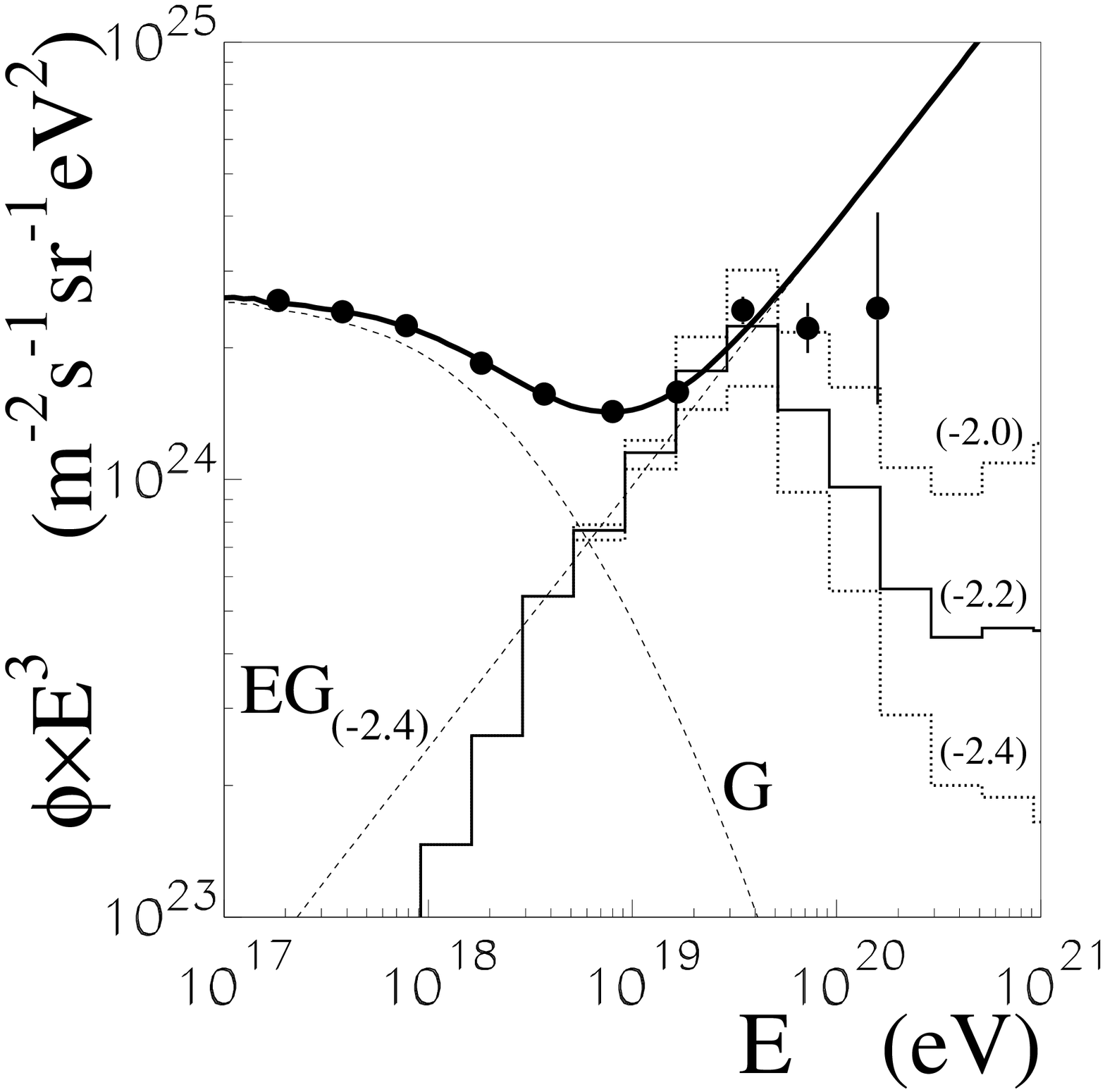}}

\noindent
Figure 1a: Primary energy spectrum of ultra-high energy cosmic rays. The points represent 
the summary of the world's data after normalization to the same 'ankle' position
and using the scale for the Hi-Res experiment \cite{9}. The sharp minimum ('ankle')
at log(E)$\sim$18.7 is regarded by us as strong evidence for a transition  from Galactic 
(G) to Extragalactic (EG) particles; primary protons are assumed in the
comparison of expectation with the points but the results for primary iron nuclei
would be similar, in view of the normalization of the expectations to the EG line
at 10$^{19}$ eV.\\
The lines represent expectations for a universal distribution of sources beyond
6 Mpc (sources closer than this would have been recognized already). The
numbers in brackets are the exponents of the injection spectra adopted in
the calculations.

\centerline{
\includegraphics[width=12.cm]{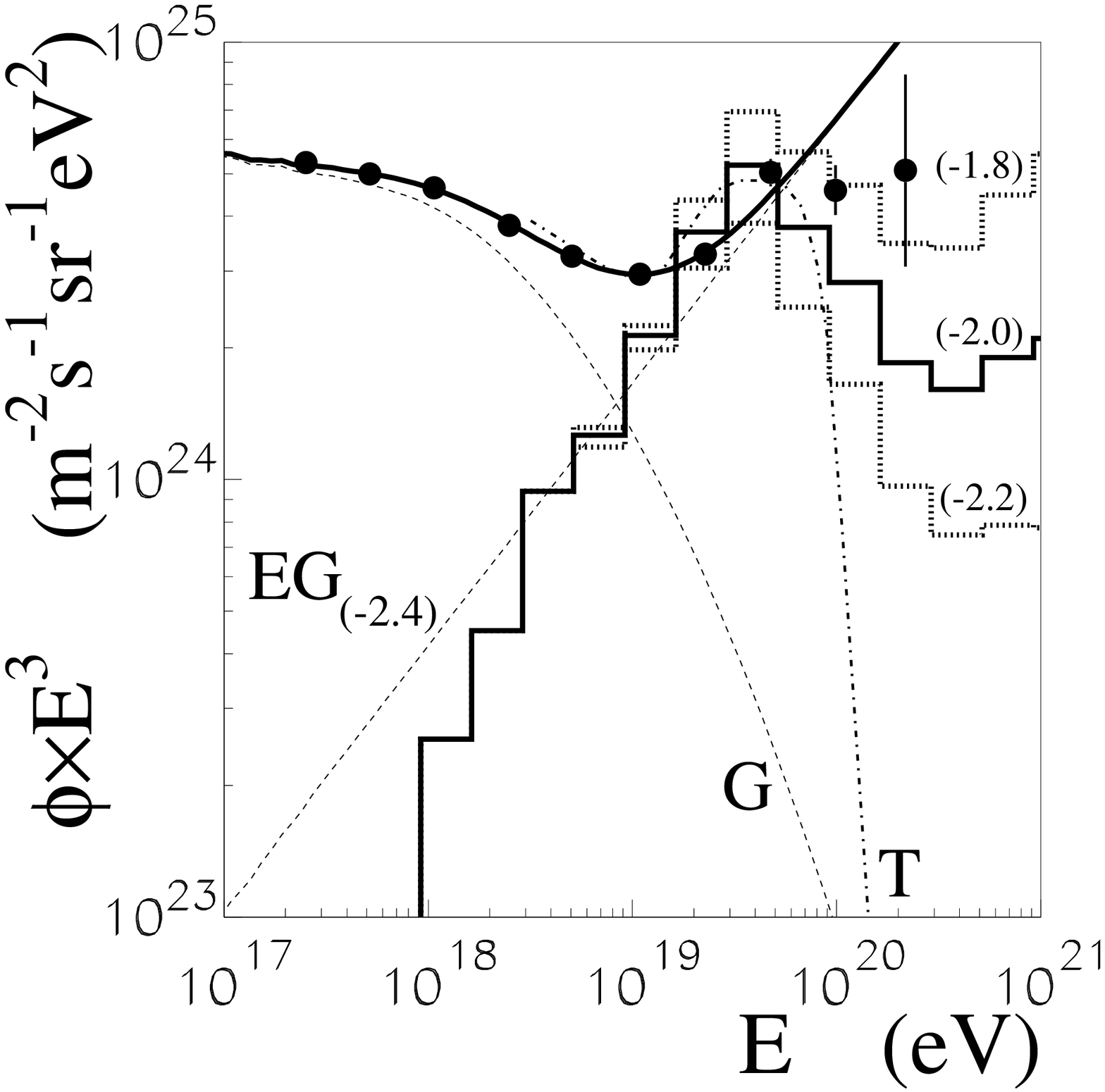}}


\noindent
Figure 1b: As figure 1(a) but for the normalization of the experimental data to the intensity and energy determined 
in the AGASA experiment \cite{9}.\\
'T' denotes a prediction commonly quoted \cite{1} but one which we regard as
inappropriate; certainly, uniform UHECR injection with an energy - independent
exponent would not give such a catastrophic fall.\\
It is evident that a GZK - 'cut-off' is neither observed nor predicted.

\centerline{
\includegraphics[width=12.cm]{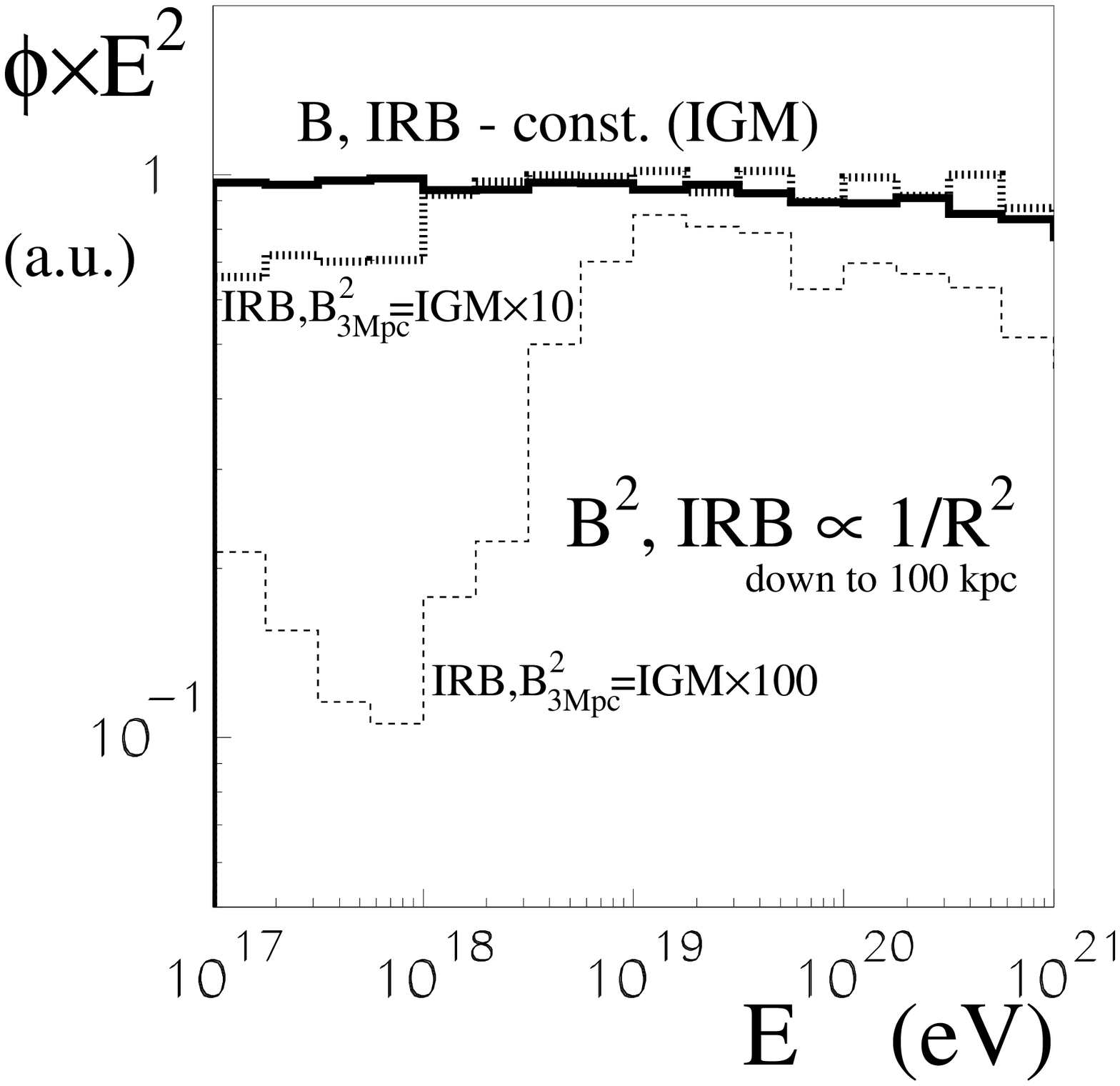}}

\noindent
Figure 2a: UHECR spectra expected on emergence from a sphere of radius 
3 Mpc round a very strong source. Various 
dependencies of the magnetic field, B, 
and infra-red intensity, IRB, on distance from the source have 
been assumed. Details are given on the graph. 
Injection spectrum of the form E$^{-2}$.\\
The actual spectra depend on the form of the cosmic ray diffusion; here we adopt the Kolmogorov formalism.\\
In calculations for models assuming injection from strong sources the assumed injection spectrum should be 
multiplied by an appropriate function of the type shown, before propagation calculations commence.

\centerline{\includegraphics[width=12.cm]{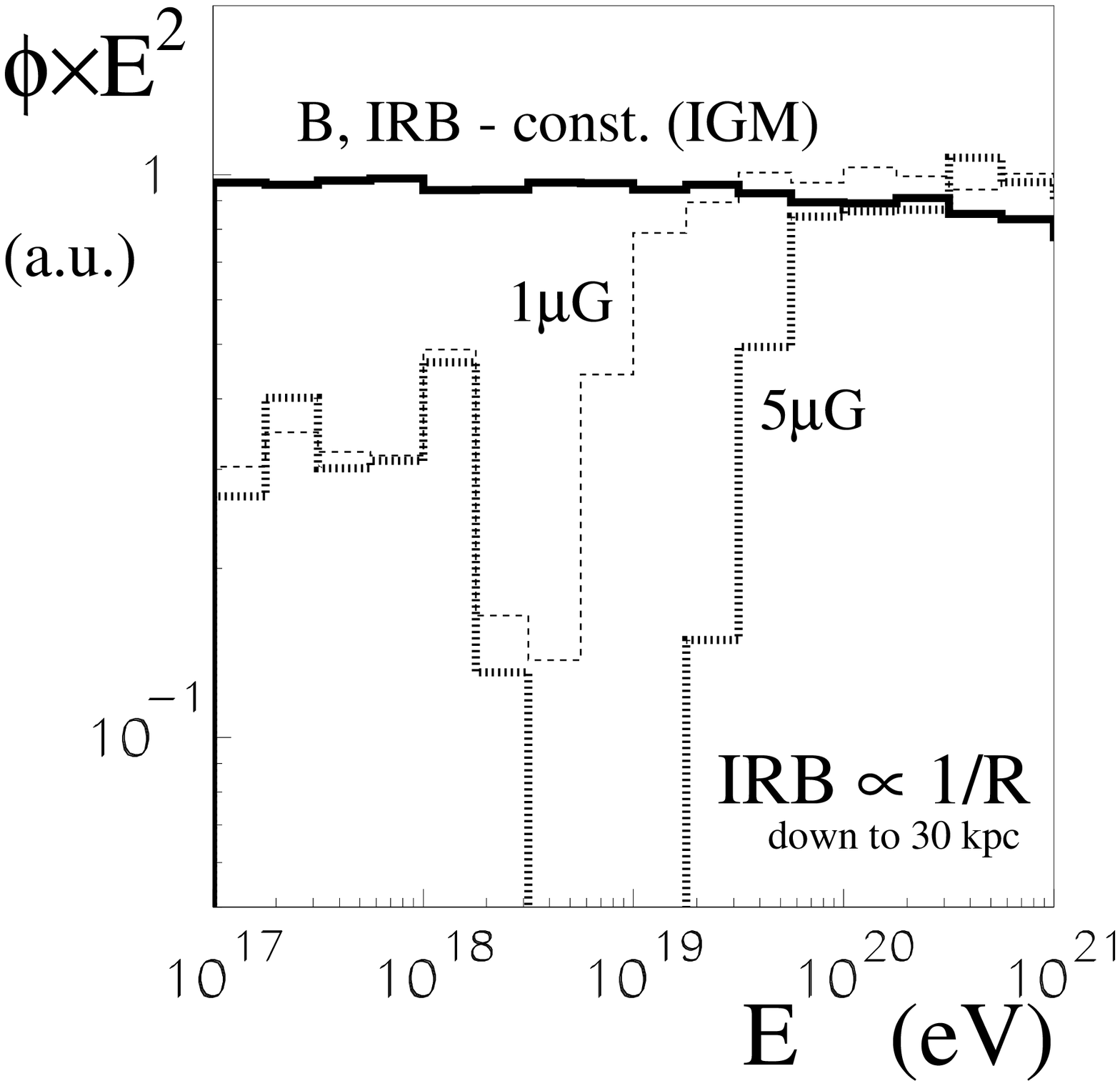}}

\noindent
Figure 2b: As figure 2(a) but for the case where the magnetic field is constant over the
cluster, of radius 3 Mpc. This field comes from emission from all the galaxies in
the cluster. A value of 5~$\mu$G is representative of a rich cluster.

\end{document}